\documentclass[reprint,twocolumn,notitlepage,preprintnumbers,nofootinbib,amsmath,amssymb,aps,floatfix,superscriptaddress,longbibliography,svgnames]{revtex4-2}

\usepackage{amsmath}
\usepackage{amsfonts}
\usepackage{amssymb}
\usepackage{graphicx}
\usepackage{bm}
\usepackage{subfigure}
\usepackage{url}
\usepackage[hyperindex]{hyperref}
\usepackage{color}
\usepackage[ddmmyy,24hr]{datetime}
\usepackage{bigdelim}
\usepackage{booktabs}
\usepackage{dcolumn}
\usepackage{multirow}
\usepackage{subfigure}
\usepackage{physics}
\usepackage{cancel}
\usepackage{stackrel}
\usepackage{paralist}
\usepackage{xspace}
\usepackage{slashed}
\usepackage{cancel}
\usepackage{todonotes}
\usepackage{enumerate}
\usepackage{float}
\usepackage{fullpage}
\usepackage{ulem}
\usepackage{wasysym}
\usepackage{comment}
\usepackage{bbold}
\usepackage{hyperref}

\usepackage{xcolor}
\usepackage{orcidlink}

\usepackage{feynmp-auto}
\newcommand{\nua}[1]{\ensuremath{\rlap{\kern-2.5pt\ensuremath{\overset{\scriptscriptstyle(-)}{\phantom{\nu}}}}{\ensuremath{{\nu}_{#1}}}}}

\usepackage{enumitem}

\begin{document}
\title{Standard Model Tested with Neutrinos}

\author{M. Atzori Corona \orcidlink{0000-0001-5092-3602}}
\email{mcorona@roma2.infn.it}
\affiliation{Istituto Nazionale di Fisica Nucleare (INFN), Sezione di Roma Tor Vergata, Via della Ricerca Scientifica, I-00133 Rome, Italy}

\author{M. Cadeddu \orcidlink{0000-0002-3974-1995}}
\email{matteo.cadeddu@ca.infn.it}
\affiliation{Istituto Nazionale di Fisica Nucleare (INFN), Sezione di Cagliari,
	Complesso Universitario di Monserrato - S.P. per Sestu Km 0.700,
	09042 Monserrato (Cagliari), Italy}

\author{N. Cargioli \orcidlink{0000-0002-6515-5850}}
\email{nicola.cargioli@ca.infn.it}
\affiliation{Istituto Nazionale di Fisica Nucleare (INFN), Sezione di Cagliari,
	Complesso Universitario di Monserrato - S.P. per Sestu Km 0.700,
	09042 Monserrato (Cagliari), Italy}

\author{F. Dordei \orcidlink{0000-0002-2571-5067}}
\email{francesca.dordei@cern.ch}
\affiliation{Istituto Nazionale di Fisica Nucleare (INFN), Sezione di Cagliari,
	Complesso Universitario di Monserrato - S.P. per Sestu Km 0.700,
	09042 Monserrato (Cagliari), Italy}

\author{C. Giunti \orcidlink{0000-0003-2281-4788}}
\email{carlo.giunti@to.infn.it}
\affiliation{Istituto Nazionale di Fisica Nucleare (INFN), Sezione di Torino, Via P. Giuria 1, I--10125 Torino, Italy}

\author{C.A. Ternes \orcidlink{0000-0002-7190-1581}}
\email{christoph.ternes@lngs.infn.it}
\affiliation{Istituto Nazionale di Fisica Nucleare (INFN),
Laboratori Nazionali del Gran Sasso, 67100 Assergi, L’Aquila (AQ), Italy}
\affiliation{Gran Sasso Science Institute, Viale F. Crispi 7, L’Aquila, 67100, Italy}

\begin{abstract}
The standard model (SM) of particle physics effectively explains most observed phenomena, though some anomalies, especially in the neutrino sector, suggest the need for extensions. In this Letter, we perform the first global fit of elastic neutrino-nucleus and neutrino-electron scattering data to further test the SM within a consistent framework. Our results on the neutrino charge radius, the only nonzero electromagnetic property of neutrinos in the SM, show no significant deviation, indicating no large beyond the SM flavor-dependent effects for electron and muon neutrinos.
By incorporating solar neutrino data from dark matter direct detection experiments, we also place the most stringent constraints on the tau neutrino charge radius obtained from neutrino scattering experiments. Additionally, we determine updated constraints on the vector and axial-vector neutrino-electron neutral current couplings, adjusting for flavor-dependent effects and for the different experimental momentum transfers. The global analysis reveals two allowed solutions: one close to the SM prediction, and a degenerate solution that is favored.
We show that future dark matter detectors could achieve sufficient precision to resolve the degeneracy. As we move toward the precision era, this Letter demonstrates the crucial need to properly account for flavor- and momentum-dependent effects to avoid misinterpretations of the data.
\end{abstract}

\maketitle  
Over time, the electroweak theory of the standard model (SM) has been extensively investigated, both theoretically and experimentally, to precisely determine the interactions between particles.
The quest for increasingly higher precision led to the introduction of the so-called radiative corrections~\cite{Marciano:2003eq,Erler:2013xha,Tomalak:2019ibg,Tomalak:2020zfh,Huang:2024rfb,Cadeddu:2019eta,Cadeddu:2018dux,AtzoriCorona:2022jeb,AtzoriCorona:2023ktl,AtzoriCorona:2024rtv}, due to higher-order vertex contributions.
Existing formalisms have been designed so that most radiative corrections in low-energy neutral current interactions are largely universal, meaning they remain independent of the specific particles participating in the process. In this Letter, we discuss neutrino interactions, focusing in particular on the elastic neutrino electron ($\nu$ES) and the coherent neutrino-nucleus scattering (CE$\nu$NS) processes, for which another nonuniversal radiative correction, the so-called neutrino charge radius (NCR), has to be accounted for. The latter, being the only nonzero neutrino electromagnetic property predicted by the SM, provides a unique way for testing the theory and looking for possible hints of flavor-dependent effects beyond the SM (BSM). Moreover, neutral current neutrino-electron couplings provide another way for testing the SM. Unfortunately, up to now the degeneracy between the vector and axial-vector neutral current couplings~\cite{ParticleDataGroup:2024cfk} has prevented attainment of a clear test of the electroweak theory. 
As a result, verifying the SM consistency has required
combination of neutrino data with $e^+e^-$ measurements.

In this Letter, we refine the conventional description of the neutrino charge-radius radiative correction by incorporating its momentum dependence~\cite{AtzoriCorona:2024rtv}. Additionally, we explicitly account for flavor- and momentum-dependent effects in the neutral current neutrino-electron couplings when combining constraints from available data. We perform the first comprehensive global analysis within a unified framework, reevaluating a vast amount of $\nu$ES and CE$\nu$NS data, to precisely test the SM using neutrinos, extending the scope of previous analyses~\cite{Barranco:2007ea,Giunti:2024gec,Hirsch:2002uv,Canas:2016vxp}.

In the SM the $\nu$ES cross section is obtained by multiplying the single electron cross section with the effective electron charge of the target atom $Z_{\text{eff}}(T_e)$~\cite{AtzoriCorona:2022qrf,Chen_2017,Kouzakov:2014lka}, and for each neutrino flavor $\nu_{\ell}$ ($\ell=e,\mu,\tau$) it is given by~\cite{ParticleDataGroup:2024cfk}
\begin{align}\label{ES-SM-flavour}
\dfrac{d\sigma_{\nu_{\ell}\,e}}{d T_{\text{e}}}&
(E,T_{e})
=
Z_{\text{eff}}(T_{e})
\,
\dfrac{G_{\text{F}}^2 m_{e}}{2\pi}
\left[
\left( g_{V}^{\nu_{\ell}\,e} + g_{A}^{\nu_{\ell}\,e} \right)^2\right.\nonumber\\\nonumber
&+
\left( g_{V}^{\nu_{\ell}\,e} - g_{A}^{\nu_{\ell}\,e} \right)^2
\left( 1 - \dfrac{T_{e}}{E} \right)^2\\
&-
\left.\left( \left(g_{V}^{\nu_{\ell}\,e}\right)^2 - \left(g_{A}^{\nu_{\ell}\,e}\right)^2 \right)
\dfrac{m_{e} T_{e}}{E^2}
\right]
,
\end{align}
where $G_{\text{F}}$ is the Fermi constant, $E$ is the neutrino energy, $m_{e}$ is the electron mass, $T_e$ is the electron recoil energy, and $g_{V}^{\nu_{\ell}\,e}$ and $g_{A}^{\nu_{\ell}\,e}$ are the neutrino-electron vector and axial-vector couplings, respectively.
To analyze the $\nu$ES process, it is necessary to study in detail the calculation of the couplings beyond tree level, thus accounting for the radiative corrections~\cite{Erler:2013xha,AtzoriCorona:2022jeb}, such as the $WW$ and $ZZ$ box diagrams as well as the NCR contribution (for more information see Supplemental Material~\cite{refsupp}). The latter enters the vector coupling and represents the only flavor-dependent radiative correction. Thus, we can define the neutrino-electron couplings as
\begin{align}
g_V^{\nu_\ell\,e}&=\tilde{g}_V^{\nu\,e}+2\diameter_{\nu_\ell W}+\delta_{\ell\, e}\, ,\label{eq:gV}\\
g_A^{\nu_\ell\,e}&=g_A^{\nu\,e}+\delta_{\ell\, e}\, ,\label{eq:gA}
\end{align}
where $\tilde{g}_V^{\nu\,e}$ and $g_A^{\nu\,e}$ are the neutral current flavor-independent couplings, and include only the flavor-independent radiative corrections. The Kronecker delta, $\delta_{\ell\, e}$, accounts for the charge current contribution, which is present only for $\ell=e$, while for antineutrinos, \(g_{A}^{\nu_\ell\, e} \to -g_{A}^{\nu_\ell\, e}\). For the full definition of the vector and axial-vector couplings please refer to Refs.~\cite{AtzoriCorona:2022jeb,refsupp}.
The remaining $\diameter_{\nu_\ell W}$ term in Eq.~(\ref{eq:gV}) represents the NCR radiative contribution, which is defined as~\cite{Erler:2013xha}
\begin{equation}
\diameter_{\nu_\ell W} = - {\alpha \over 6 \pi} \left( \ln {M_W^2 \over m_\ell^2} + {3 \over 2} \right),\label{phiNCR}
\end{equation}
at zero-momentum transfer. Here, $\alpha$ is the low-energy limit of the electromagnetic coupling, $M_W$ is the $W$ boson mass and $m_\ell$ is the mass of the charged lepton with flavor $\ell$. Clearly, this radiative contribution depends on the neutrino flavor and is generated by a loop insertion into the $\nu_\ell$ line, where $W$ boson(s) and charged lepton(s) $\ell$ enter. These diagrams' contribution can be calculated, and according to Refs.~\cite{Bernabeu:2000hf,Bernabeu:2002nw,Bernabeu:2002pd}, the NCR corresponds to a physical observable, being finite and gauge invariant. In particular, the SM calculation renders~\cite{Cadeddu:2018dux}
\begin{equation}
    \langle r^2_{\nu_\ell}\rangle_{\rm SM}=-\dfrac{G_F}{2\sqrt{2}\pi^2}\Big[3-2\ln \Big(\dfrac{m_\ell^2}{M_W^2}\Big)\Big]\, ,   \label{NCRdef}
\end{equation}
which corresponds to the numerical values
\begin{align}
    \langle r^2_{\nu_e}\rangle &\simeq-0.83\times 10^{-32}\, \mathrm{cm}^2\, ,\label{NCReeSM}\\
    \langle r^2_{\nu_\mu}\rangle &\simeq-0.48\times 10^{-32}\, \mathrm{cm}^2\, ,\label{NCRuuSM}\\
    \langle r^2_{\nu_\tau}\rangle &\simeq-0.30\times 10^{-32}\, \mathrm{cm}^2\, .\label{NCRttSM}
\end{align}
The numerical values for the neutrino-electron couplings in the low-energy limit are
\begin{align}
   g_{V}^{\nu_{e}\,e} &= 0.9524, \quad g_{A}^{\nu_{e}\,e} = 0.4938, \label{eq:gve}\\
   g_{V}^{\nu_{\mu}\,e} &= -0.0394, \quad  g_{A}^{\nu_{\mu}\,e} = -0.5062, \label{eq:gvm}\\
   g_{V}^{\nu_{\tau}\,e} &= -0.0350, \quad g_{A}^{\nu_{\tau}\,e} = -0.5062,  
   \label{eq:gvt}
\end{align}
when including radiative corrections~\cite{AtzoriCorona:2022jeb,Erler:2013xha} and the latest weak mixing angle calculation~\cite{ParticleDataGroup:2024cfk}. 

Given that the NCR contributes not only to the $\nu$ES process, but also to CE$\nu$NS by entering into the neutrino-proton coupling (since it affects the scattering of neutrinos with charged particles~\cite{AtzoriCorona:2024rtv}) the complementarity between $\nu$ES and CE$\nu$NS is pivotal to fully leveraging the data.

Although measuring the NCR would serve as a fundamental test of the SM, the available data remain insufficient to provide a first determination. So far, only constraints have been put on its value~\cite{AtzoriCorona:2022qrf,DeRomeri:2022twg,TEXONO:2009knm,Ahrens:1990fp,A:2022acy}. When considering BSM effects on the NCR, one should also consider possible off-diagonal flavor-changing contributions, known as transition charge radii~\cite{Cadeddu:2018dux}. Constraints on these have been derived from existing data (see, e.g., Refs.~\cite{AtzoriCorona:2022qrf,Giunti:2023yha}). Here, however, we focus solely on the diagonal NCRs, which allows us to probe their SM values.
 \begin{figure*}[t!]
    \centering
    \includegraphics[width=0.49\linewidth]{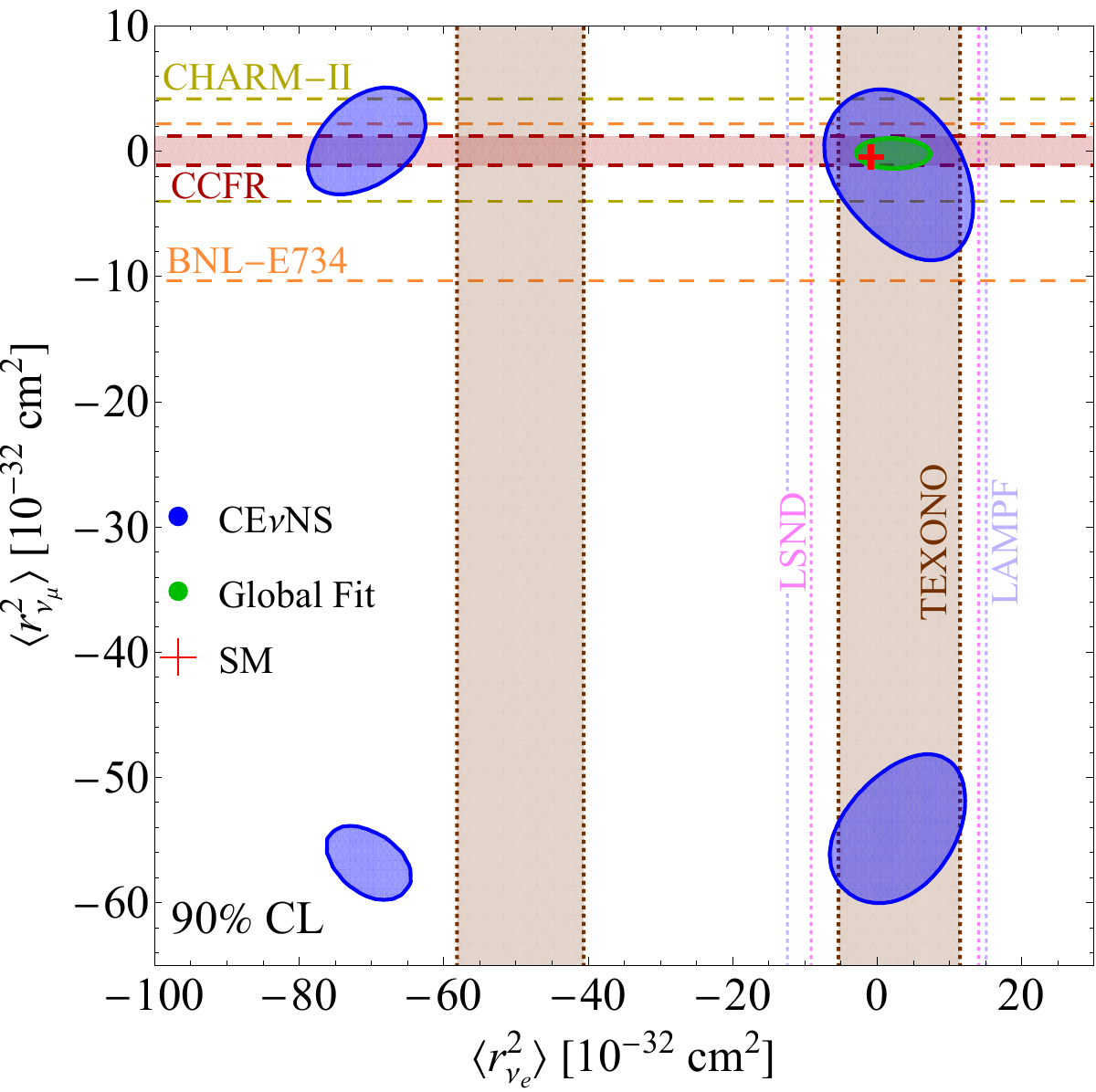}
    \includegraphics[width=0.5\linewidth]{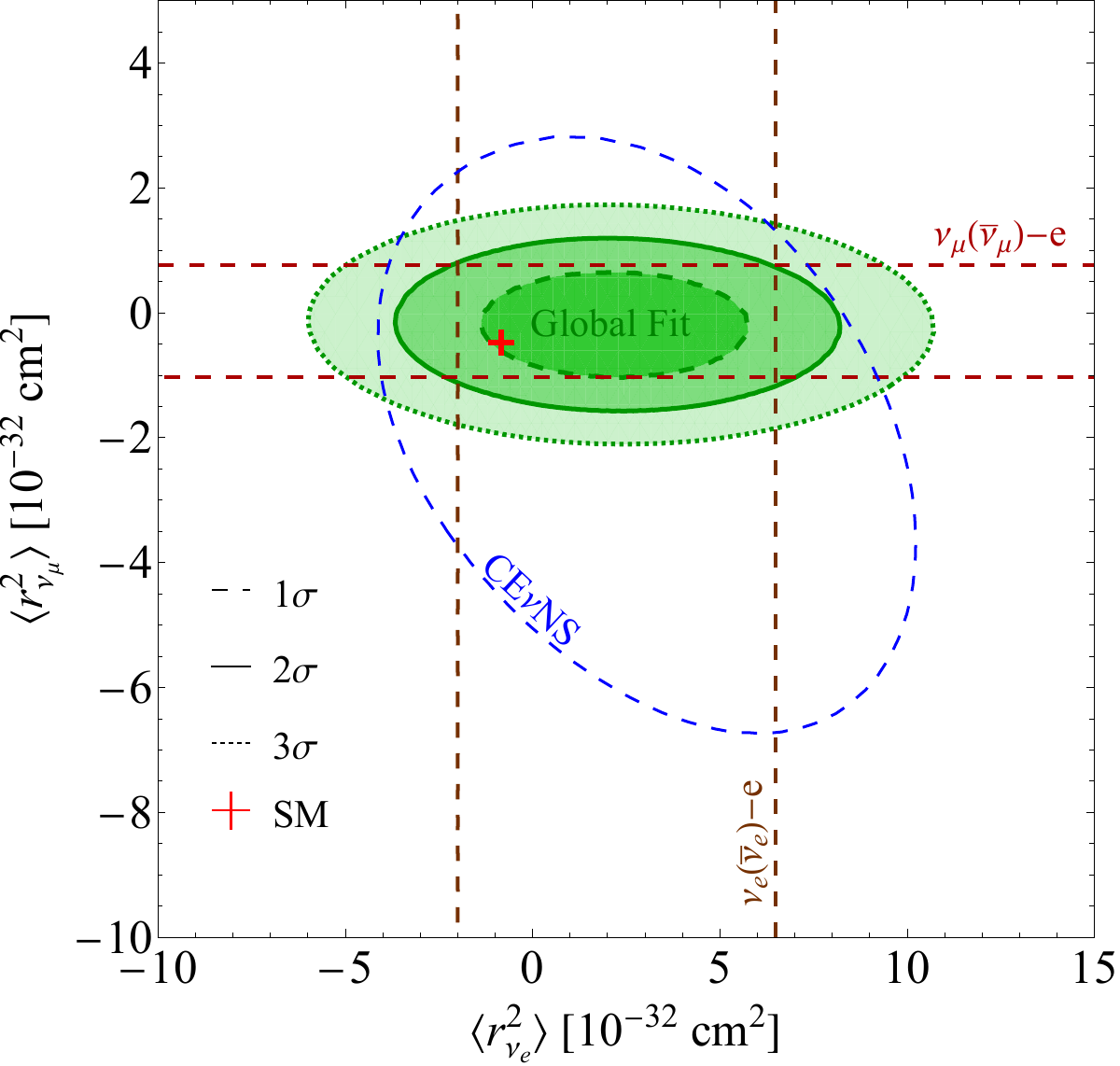}
    \caption{Left: contours at 90\% CL for 2 degrees of freedom ($\Delta\chi^2\simeq4.61$) obtained from the analysis of $\nu$ES data from TEXONO~\cite{TEXONO:2009knm}, LSND~\cite{LSND:2001akn}, LAMPF~\cite{Allen:1992qe}, CHARM-II~\cite{CHARM-II:1994dzw,CHARM-II:1994aeb}, BNL-E734~\cite{Ahrens:1990fp}, and CCFR~\cite{CCFR:1997zzq} along with the constraints obtained by combining the available CE$\nu$NS data~\cite{AtzoriCorona:2025ygn,AtzoriCorona:2024rtv} with the addition of $\nu$GEN data~\cite{nuGeN:2025mla}, and their combination (green contour). The red cross indicates the SM value of the NCRs. Right: the same constraints shown around the SM prediction and at different CLs. The dashed vertical band indicates the combined $\nu_e-e$ and $\overline{\nu}_e-e$ result, while the horizontal one the combined $\nu_\mu-e$ and $\overline{\nu}_\mu-e$ one at 1$\sigma$ CL for 2 degrees of freedom ($\Delta\chi^2\simeq2.30$).}
    \label{fig:2DNCR}
\end{figure*}

As discussed recently in Ref.~\cite{AtzoriCorona:2024rtv}, the neutrino charge radius radiative correction depends also on the momentum transfer, which therefore should be carefully corrected for. For this purpose, we include a ``neutrino charge radius form factor", which is defined by isolating the momentum-dependent NCR with respect to the SM picture, so basically as
\begin{equation}
    \mathcal{F}_{\nu_\ell}(q^2)=\dfrac{\langle r^2_{\nu_\ell}\rangle^{\rm eff}(q^2)}{\langle r^2_{\nu_\ell}\rangle^{\rm eff}(0)}\, \equiv \dfrac{\langle r^2_{\nu_\ell}\rangle^{\rm eff}(q^2)}{\langle r^2_{\nu_\ell}\rangle^{\rm SM}}\, ,\label{FFncr}
\end{equation}
where we introduced an effective NCR definition~\cite{AtzoriCorona:2024rtv}. The neutrino-electron vector neutral coupling can then be defined as
\begin{equation}
g_V^{\nu_\ell\,e}=\tilde{g}_V^{\nu\,e}+\dfrac{\sqrt{2}\pi\alpha}{3 G_{\text{F}}}\langle{r^2_{\nu_\ell}}\rangle \mathcal{F}_{\nu_\ell}(q^2)+\delta_{\ell\, e}\, \label{vectorCoup}.
\end{equation} 
The impact of the momentum dependence of the NCR form factor becomes relevant for momenta larger than the mass of the corresponding charged lepton $\ell$ that enters the loops.

We can profit from the large amount of available $\nu$ES data to extract the value of the NCRs and combine them with the constraints set by us using the available CE$\nu$NS data in Ref.~\cite{AtzoriCorona:2025ygn}, namely from COHERENT CsI~\cite{Akimov:2021dab}, COHERENT Ar~\cite{COHERENT:2020iec,COHERENT:2020ybo}, CONUS+~\cite{Ackermann:2025obx} and TEXONO Ge~\cite{TEXONO:2024vfk}. For completeness, we also updated such constraints including the latest $\nu$GEN~\cite{nuGeN:2025mla} data. Moreover, in this Letter, we analyze the reactor $\overline{\nu}_e-e$ data from the TEXONO collaboration~\cite{TEXONO:2009knm} using a recent antineutrino flux parametrization~\cite{Perisse:2023efm}. We also consider the results on the $\nu_e-e$ integrated cross section measured by the LSND~\cite{LSND:2001akn} and LAMPF~\cite{Allen:1992qe} collaborations.
Similarly, we analyze the integrated cross section results reported by BNL-E734 for the $\nu_\mu (\overline{\nu}_\mu)-e$ processes~\cite{Ahrens:1990fp}.
We also include the $\nu_\mu (\overline{\nu}_\mu)-e$ differential cross section measurement reported by the CHARM-II experiment~\cite{CHARM-II:1993phx,CHARM-II:1994dzw,CHARM-II:1994aeb}. Lastly, to incorporate also the data from the CCFR collaboration~\cite{CCFR:1997zzq}, we converted the constraint on the weak mixing angle into a determination of the muonic NCR, which can be done given the fact that the neutrino charge radius acts as an effective shift of the weak mixing angle inside the vector neutrino-electron neutral current coupling (e.g., see Eq.~(1) of Ref.~\cite{Cadeddu:2018dux}).

The strength of this Letter lies in the development of a global fit that, for the first time, incorporates all the most relevant $\nu$ES and CE$\nu$NS data within a consistent theoretical and phenomenological framework.
Specifically, we build upon previous analyses which either neglected radiative corrections altogether or included them only partially, typically considering only the neutrino charge radius correction. In contrast, our Letter incorporates the full set of relevant radiative contributions (as detailed in Supplemental Material~\cite{refsupp}), applies a consistent treatment across all experimental datasets, ensuring uniformity across experiments and allowing for a more accurate and meaningful comparison of different data, uses the most up-to-date constants and a consistent antineutrino flux description for reactor data, and, most importantly, includes the momentum dependence that was pioneered in Ref.~\cite{AtzoriCorona:2024rtv}.

The results from our analysis of the individual aforementioned data are shown in Fig.~\ref{fig:2DNCR} for the electron and muon neutrino charge radii, along with their combination, obtained by keeping $\tilde{g}_V^{\nu\,e}$ fixed to its SM value and varying $\langle{r^2_{\nu_\ell}}\rangle$.
The $1\sigma$ confidence level (CL) NCR values obtained from the global fit are
\begin{align}
    \langle{r_{\nu_{e}}^{2}}\rangle=&2.2^{+2.4}_{-2.3}\,\times 10^{-32}\, \mathrm{cm}^2\, ,
    \label{eq:CR_limit_ee}\\ 
    \langle{r_{\nu_{\mu}}^{2}}\rangle=&-0.19^{+0.55}_{-0.56}\,\times 10^{-32}\, \mathrm{cm}^2\, ,
    \label{eq:CR_limit_uu}
\end{align}
which are fairly in agreement with the SM prediction. These represent the most precise determinations of such quantities.
Note that none of the individual datasets is able to select the SM region alone\footnote{In the literature, most of the analyses focused on the SM region itself, ignoring the existence of other degenerate solutions. See supplemental material~\cite{refsupp} for more information.}, which only emerges through the global fit.
\begin{figure}[t]
    \centering
    \includegraphics[width=\linewidth]{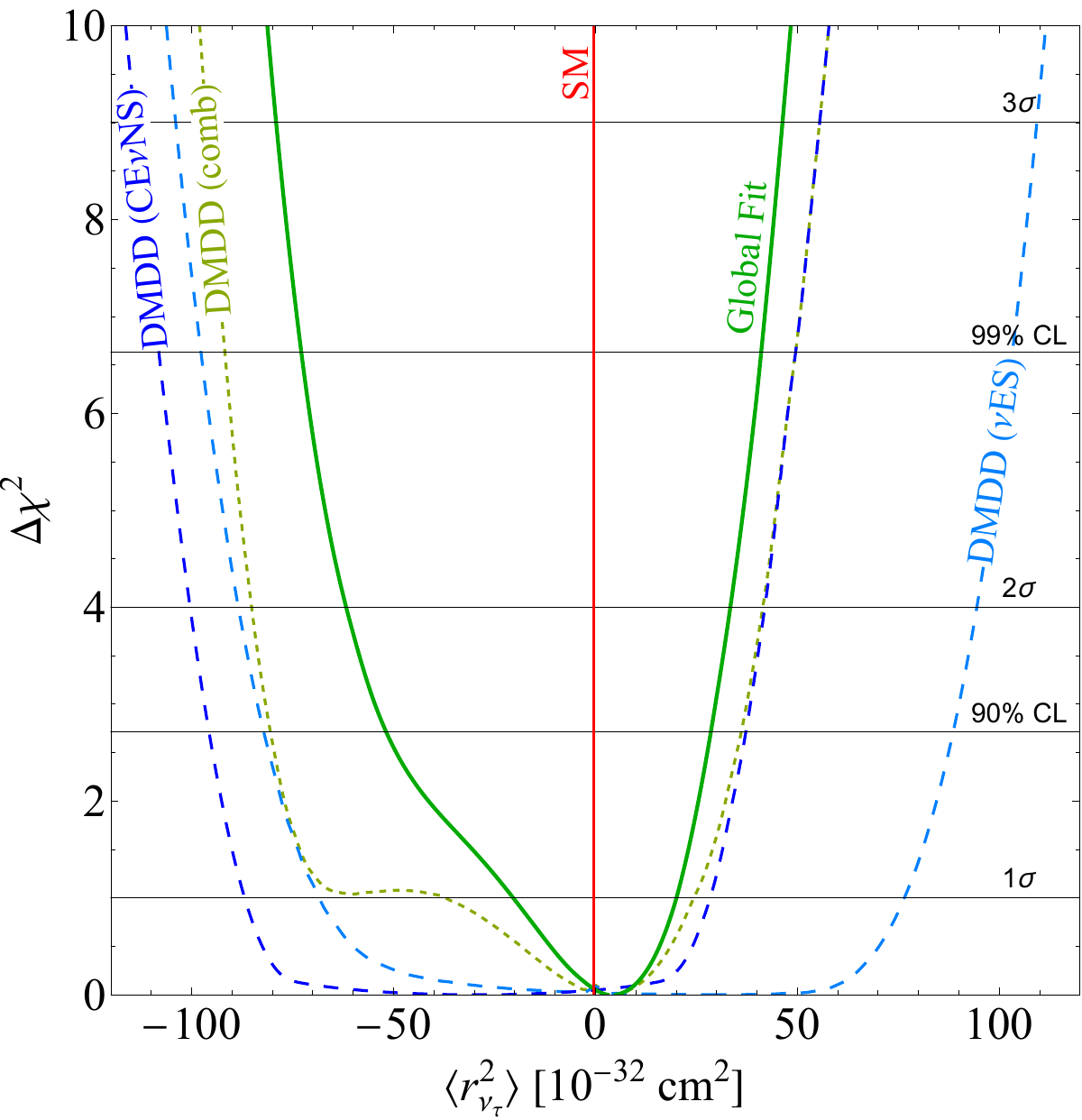}
    \caption{Marginal $\Delta\chi^2$'s obtained from the analysis of direct detection dark matter experiments (DMDDs)~\cite{DeRomeri:2024iaw,DeRomeri:2024hvc,Giunti:2023yha,AtzoriCorona:2022jeb} analyzing the signals due to solar neutrino CE$\nu$NS and $\nu$ES. The solid green curve indicates the result from the global fit.}
    \label{fig:1DNCRsolar}
\end{figure}

While putting less precise constraints on $\langle r^2_{\nu_{e}}\rangle$ and $\langle r^2_{\nu_{\mu}}\rangle$, solar neutrino data from dark matter direct detection experiments (DMDD) are uniquely sensitive also to the $\tau$ flavor both through CE$\nu$NS~\cite{XENON:2024ijk,PandaX:2024muv} and $\nu$ES~\cite{PandaX:2022ood,LZ:2022lsv,XENON:2022ltv}. Including solar neutrino data in the global analysis yields the most stringent constraint on the $\tau$ flavor obtained from CE$\nu$NS and $\nu$ES experiments. For the solar neutrino analysis we closely follow Refs.~\cite{DeRomeri:2024iaw,DeRomeri:2024hvc} for CE$\nu$NS data and Refs.~\cite{Giunti:2023yha,AtzoriCorona:2022jeb} for the $\nu$ES case. The results from the several analyses are shown in Fig.~\ref{fig:1DNCRsolar}. The $1\sigma$ CL from the global fit corresponds to 
\begin{align}
    -20\leq\langle{r_{\nu_{\tau}}^{2}}\rangle [10^{-32}\, \mathrm{cm}^2]\leq20 \,.
    \label{eq:CR_limit_tt}
\end{align}
Note that even though the previously discussed experiments do not measure $\langle r^2_{\nu_\tau} \rangle$ directly, their inclusion helps to break degeneracies, improving the overall bound on the $\tau$ flavor as shown in the figure. The bound is still less precise than the ones for the other neutrino flavors, although excellent prospects are awaited considering next-generation experiments~\cite{Giunti:2023yha}. 

Restricting ourselves only to the $\nu$ES data, we can constrain the value of the vector and axial-vector neutrino-electron neutral current couplings to test the prediction of the electroweak theory. In the literature, the constraints are obtained by directly fitting $g_V^{\nu_\ell\,e}$ and $g_A^{\nu_\ell\,e}$ inside the neutrino-electron cross section. Due to the radiative corrections and the charge current contribution, the vector coupling naturally depends on the neutrino flavor, as shown in Eq.~(\ref{eq:gV}). Thus, to correctly compare the available data, we restrict our analysis to the flavor-independent neutral current couplings, $\tilde{g}_V^{\nu\,e}$ and  $\tilde{g}_A^{\nu\,e}$, while keeping the neutrino charge radius contributions fixed to their SM predictions.
However, $\tilde{g}_V^{\nu\,e}$ also depends on the experimental energy scale, e.g. due to the effect of the running of the weak mixing angle, which enters only the vector coupling. To combine correctly the different measurements, we thus correct for the dependence on the energy scale, and compare the value of the couplings at a common scale, chosen to be the $Z$ boson mass, $M_Z$, as often done in the context of the weak mixing angle~\cite{ParticleDataGroup:2024cfk}.
\begin{figure*}[t!]
    \centering
    \includegraphics[width=0.49\linewidth]{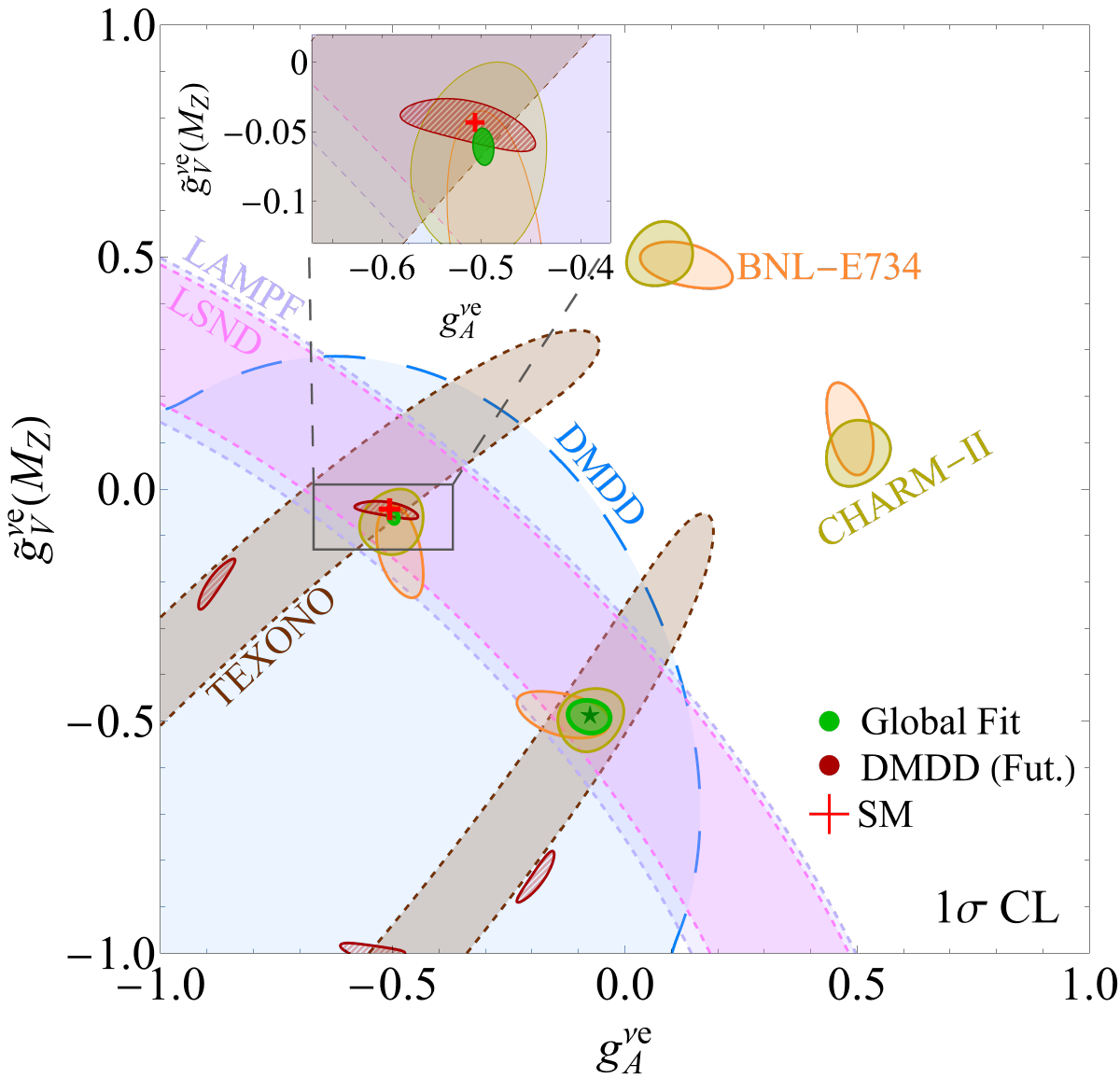}
    \includegraphics[width=0.49\linewidth]{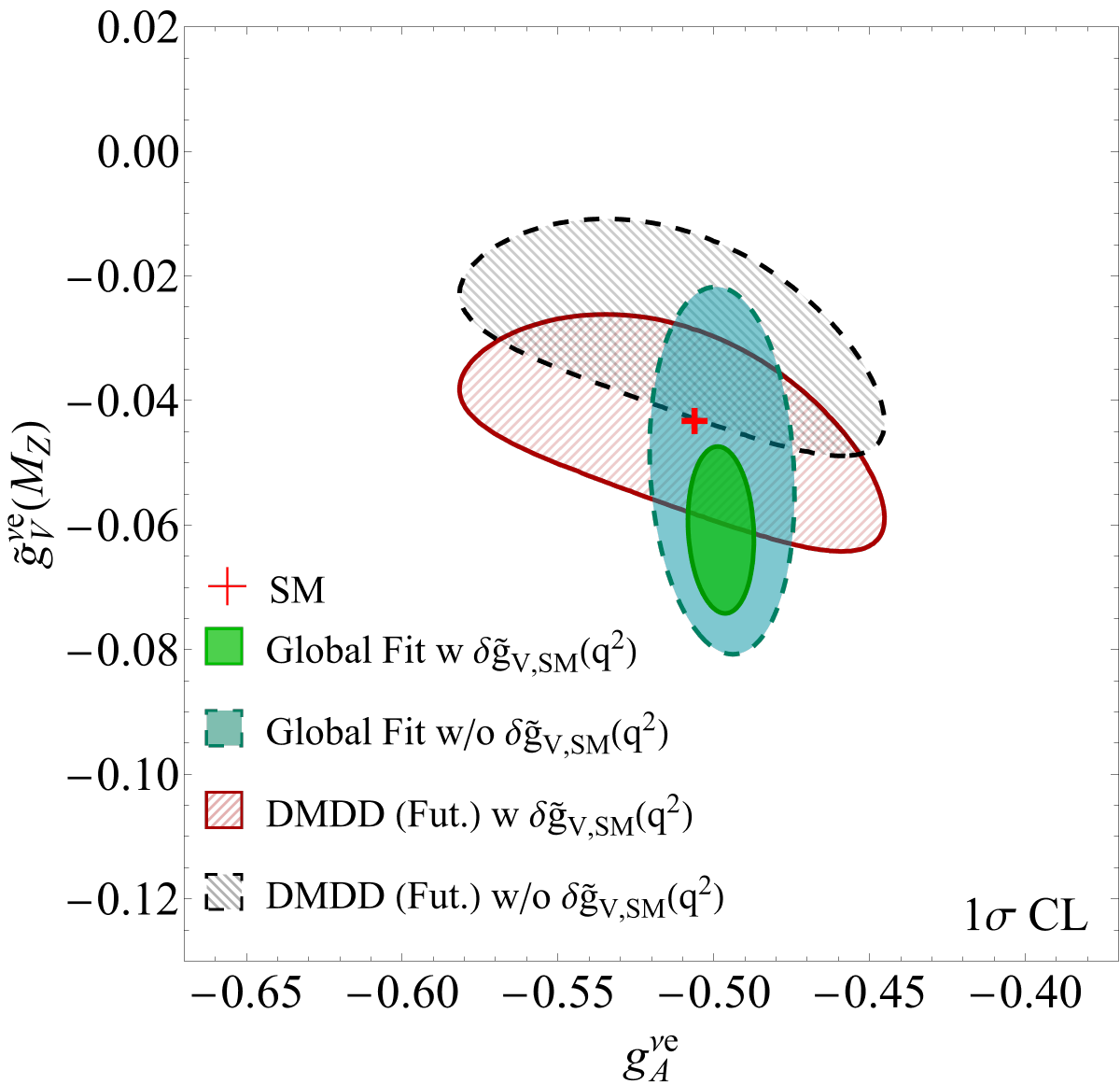}
    \caption{Left: constraints at $1\sigma$ CL on the flavor-independent neutral current couplings $\tilde{g}_V^{\nu\,e}(M_Z)$ and the $g_A^{\nu\,e}$ from the different $\nu$ES data along with their combination. The red cross indicates the SM prediction while the green star indicates the best fit. The dark red striped contour shows the potentiality of a future DMDD experiment~\cite{DARWIN:2020bnc}. In the inset, we show an enlargement of the contours around the SM prediction. 
    Right: comparison of the global fit results with the sensitivity from a future DMDD experiment, including (solid) or not (dashed) the correction due to the momentum transfer.}
    \label{fig:2Dgvga}
\end{figure*}

To this aim, we introduce a term, $\delta \tilde{g}_V$, in the definition of the coupling in Eq.~(\ref{vectorCoup}) which accounts for the running of the weak mixing angle as predicted by the SM. Namely, $\delta \tilde{g}_V$ has been defined as the difference between the coupling constant at a certain energy scale $q^2$ with respect to its value at $M_Z$
\begin{equation}
    \delta \tilde{g}_{V,\, \mathrm{SM}}(q^2)=\tilde{g}_{V,\, \mathrm{SM}}^{\nu\,e}(q^2)-\tilde{g}_{V,\, \mathrm{SM}}^{\nu\,e}(M_Z)\, ,
\end{equation}
so that the flavor-independent coupling at a generic scale $q^2$ can be written in terms of its value at $M_Z$
\begin{equation}
    \tilde{g}_V^{\nu\,e}(q^2)=\tilde{g}_V^{\nu\,e}(M_Z)+\delta\tilde{g}_{V,\, \mathrm{SM}}(q^2)\, .
\end{equation}
By substituting the above definition inside Eq.~(\ref{vectorCoup}), we can measure directly $\tilde{g}_V^{\nu\,e}(M_Z)$, which in this way is a universal quantity independent of the particular experiment.
The correction $\delta\tilde{g}_{V,\, \mathrm{SM}}^{\nu\,e}(q^2)$ has a larger impact for low energy data. Indeed, $\delta\tilde{g}_{V,\, \mathrm{SM}}^{\nu\,e}(q^2=0)$ is about 50\% of $\tilde{g}_{V,\, \mathrm{SM}}^{\nu\,e}(q^2=0)$.

In Fig.~\ref{fig:2Dgvga} (left), we show the results at 1$\sigma$ CL on the neutral current couplings obtained accounting for such a momentum dependence in the considered $\nu$ES experiments along with their combination. Differently from the results reported by the latest PDG electroweak review, in Fig.~10.1 of  Ref.~\cite{ParticleDataGroup:2024cfk}, we separate the flavor-dependent contribution due to the NCR and correct for the different experimental momentum transfer, to extract the neutral current flavor-independent coupling evaluated at $M_Z$.
The global fit indicates a preference for the degenerate solution (obtained by swapping $\tilde{g}_{V,\, \mathrm{SM}}^{\nu\,e}$ and $g_{A,\, \mathrm{SM}}^{\nu\,e}$), while the SM prediction, as visible in the inset, lies just outside the $1\sigma$ contour. The difference in chi square between the two solutions is about $\Delta\chi^2\simeq 2.1$, while the allowed values at 90\% CL are
\begin{align}   
\tilde{g}_V^{\nu\,e}(M_Z)=[-0.53,-0.45]&\cup[-0.08,-0.03]\, ,\label{eq:gV_limit}\\
g_A^{\nu\,e}=[-0.52,-0.48]&\cup[-0.13,-0.03]\, ,\label{eq:gA_limit}
\end{align}
to be compared with the SM values ${\tilde{g}_{V,\, \mathrm{SM}}^{\nu\,e}(M_Z)=-0.0433}$ and $g_{A,\, \mathrm{SM}}^{\nu\,e}=-0.5062$. 

If such a preference persists, new and more precise $\nu$ES measurements would be needed.
In Fig.~\ref{fig:2Dgvga} (left), we overlay the obtained contours with the sensitivity expected from a future xenon-based DMDD experiment with an exposure of 300 ton years~\cite{DARWIN:2020bnc,Giunti:2023yha}. 
It is interesting to notice that the expected precision will reach the level of our current global fit, thus
showing the capability of discriminating the SM solution with respect to the degenerate one, without the need of including $e^+e^-$ data with additional assumptions~\cite{ParticleDataGroup:2024cfk}. Thus, future DMDD experiments will play a crucial role in testing the standard model with neutrinos only. However, such a level of precision requires one to carefully account for the effects investigated in this Letter. In fact, even if the effect of the momentum transfer plays a small but nonnegligible role in the global fit of the current data, in the future it will be mandatory to account for it. This is shown in Fig.~\ref{fig:2Dgvga} (right), where, restricting ourselves around the SM values for the couplings, the result of the global fit performed by correcting for the momentum transfer is compared to the case in which no correction is included. We superimposed the contours obtained by the sensitivity study of a future DMDD in the same two scenarios. 
Neglecting the effect of momentum transfer results in a significant shift, as large as 50\%, on $\tilde{g}_{V}^{\nu\,e}(M_Z)$.
Future proposed measurements with high precision such as DUNE~\cite{deGouvea:2019wav}, future ultranear reactor experiments~\cite{Brdar:2024lud}, and the LHC Forward Physics Facility~\cite{MammenAbraham:2023psg} will provide similar crucial information for testing the SM with neutrinos.\\ \\
To conclude, in this Letter we present the state-of-the-art of a global fit of neutrino data, in particular reanalyzing a vast sample of neutrino-electron and neutrino-nucleus scattering data, to extract the most stringent constraints on the neutrino charge radius, which show no significant evidence for flavor-dependent deviations from the SM picture. Moreover, we set stringent constraints on the vector and axial-vector neutral current couplings to perform a robust test of the electroweak theory. The available data are not sufficient to exclude the degenerate solution, which remains the preferred one. Thus, we emphasize the importance of radiative corrections and momentum-dependent effects, highlighting their critical role in future precision measurements.

\bibliographystyle{apsrev4-2}
\bibliography{ref}

\end{document}


\appendix

\begin{center}
 \bf \large Supplemental Material
\end{center}
\vspace*{0.2cm}

In the following appendices we provide additional details about the results of our global fit. 
In Appendix~\ref{app:rad} we describe the calculation of the neutrino-electron couplings, taking into account the radiative corrections; in Appendix~\ref{app:comparison} we compare our cross section evaluation with those available in the literature;
in Appendix~\ref{app:NCR} we discuss the constraints on the neutrino charge radii and in Appendix~\ref{app:gVgA} we enter in more detail about the effect of the experimental momentum transfer on the flavor independent neutral current neutrino-electron coupling which arises directly from the running of the weak mixing angle.

\section{Radiative corrections to $\nu$ES process}
\label{app:rad}

Here, we report details about the calculation of the neutrino-electron couplings beyond the tree-level, thus taking into account the radiative corrections~\cite{Erler:2013xha,AtzoriCorona:2022jeb,AtzoriCorona:2024rtv}.
In particular, vector and axial-vector $\nu_\ell$-electron couplings  are given by 
\begin{align}\nonumber
g_V^{\nu_\ell\,e}&=\tilde{g}_{V}^{\nu\, e}+2\diameter_{\nu_\ell W}+\delta_{\ell\,e}=\nonumber\\
&=\rho\left(-\frac{1}{2}+2 s^2_0\right)+\Box_{WW}+\rho(\boxtimes_{ZZ}^{eL}-\boxtimes_{ZZ}^{eR})+\nonumber\\
&+2\diameter_{\nu_\ell W}+\delta_{\ell\,e},\label{eq:appgv}\\
g_A^{\nu_\ell\,e}&=g_{A}^{\nu\, e}+\delta_{\ell\,e}=\nonumber\\
&=\rho\left(-\frac{1}{2}+\boxtimes_{ZZ}^{eL}+\boxtimes_{ZZ}^{eR}\right)+\Box_{WW}+\delta_{\ell\,e},\label{eq:appga}
\end{align}
where $L,R$ indicate left- and right-handed electrons, respectively, and $\delta_{\ell\,e}$ accounts for the charge current contribution to the electron-neutrino couplings.
In these relations, $\rho= 1.00063$ represents a low-energy correction for neutral-current processes while \mbox{$s_0^2\equiv\sin^2{\theta}_W(q^2=0)=0.23873\pm0.00005$}~\cite{ParticleDataGroup:2024cfk} indicates the low-energy value of the weak mixing angle predicted within the SM. The other corrections inserted come from different contributions, such as the neutrino charge radius ($\diameter_{\nu_\ell W}$) and EW box diagrams ($\boxtimes_{ZZ}^{f X}$, $\square_{WW}$).
They can be expressed as
\begin{align}
\label{WWbox}
& \diameter_{\nu_\ell W} = - {\alpha \over 6 \pi} \left( \ln {M_W^2 \over m_\ell^2} + {3 \over 2} \right),\\
&\Box_{WW} = - {\hat\alpha_Z \over 2 \pi \hat s_Z^2}\left[ 1 - {\hat\alpha_s(M_W) \over 2 \pi} \right], 
\end{align}
\be\label{eq:ZZbox}
\boxtimes_{ZZ}^{fX} = - {3 \hat\alpha_Z \over 8 \pi s_Z^2 \hat c_Z^2} (g_{LX}^{\nu_\ell f})^2
\left[ 1 - {\hat\alpha_s(M_Z) \over\pi} \right],
\ee
where $X\in\{L,R\}$, 
$\hat{\alpha}_Z\equiv \alpha(M_Z)$ is the fine structure constant at the $Z$ boson mass and $\hat{\alpha}_s(M_{W(Z)})$ is the strong constant at the $W(Z)$ boson mass.
Note that, in Eq.~(\ref{eq:ZZbox}) the $(g_{LX})^{\nu_\ell f}$ are given by $g_{LL}^{\nu_\ell e} =-\frac{1}{2}+s^2_Z$ and $g_{LR}^{\nu_\ell e} = s^2_Z$, where $s^2_Z$ is the weak mixing angle evaluated at the $Z$ mass, namely $s^2_Z=0.23129$~\cite{ParticleDataGroup:2024cfk}.
For experiments at higher energies, we can generalize the $g_V^{\nu_\ell\,e}$ coupling by substituting the low-energy value of the weak mixing angle, $s^2_0$, with its value at the proper energy scale as predicted by the running of the weak mixing angle, $\sin^2\theta_W(q^2)$. In this sense, $\tilde{g}_{V}^{\nu\, e}\equiv\tilde{g}_{V}^{\nu\, e}(q^2)$. Moreover, the $\diameter_{\nu_\ell W}$ has to be replaced with its generalized definition, as described in detail in Ref.~\cite{AtzoriCorona:2024rtv},
\begin{align}
    &\diameter_{\nu_\ell W}^\mathrm{eff}(q^2)=-\dfrac{\alpha}{\pi}\left(-R_\ell(q^2)+\dfrac{1}{4}\right) \\
    &=-\dfrac{\alpha}{\pi}\left(-\int_0^1dx\, x(1-x)\ln\Big[\dfrac{m_\ell^2-q^2x(1-x)}{M_W^2}\Big]+\dfrac{1}{4}\right)\, ,\nonumber
\end{align}
which can be translated into an effective definition of the neutrino charge radius, 
\begin{equation}
    \langle r_{\nu_\ell}^2\rangle^{\rm eff}=\dfrac{6\, G_F}{\sqrt{2}\pi\alpha}\diameter^{\rm eff}_{\nu_\ell W}(q^2)\, .
\end{equation}

\section{Comparison with literature}
\label{app:comparison}
In the literature, other works have discussed radiative corrections in the context of neutrino scattering processes adopting different frameworks~\cite{Tomalak:2020zfh,Tomalak:2019ibg,Miranda:2021mqb,Huang:2024rfb,SEHGAL1985370}. The results reported in this work are in good agreement with these calculations. In particular, in Ref.~\cite{Tomalak:2020zfh} the authors discuss the flavor dependence of radiative contributions within an effective field theory framework, including also the effects of momentum transfer on the neutrino charge radius contribution.
In Tab.~\ref{tab:cs-comparison} we report our numerical values for the CE$\nu$NS cross section on argon (Ar) for three different neutrino energies. These results can be directly compared to the values reported in Tab.~2 of Ref.~\cite{Tomalak:2020zfh}, showing an excellent agreement within less than 0.1\%, which is well below the current experimental precision.
\begin{table}[h]
    \centering
    \renewcommand{\arraystretch}{1.2}
    \resizebox{0.85\columnwidth}{!}{
    \begin{tabular}{c|c|c}
         $E_\nu\, [\mathrm{MeV}]$ & $\sigma^{\rm TL}_{\nu_\mu-\text{Ar}} \,[10^{-40}\mathrm{cm^2}]$ &  $\sigma^{\rm rad}_{\nu_\mu-\text{Ar}} \,[10^{-40}\mathrm{cm^2}]$  \\\hline
         10 &  1.77 &  1.91 \\\hline
         30  & 14.24 &  15.37  \\\hline
         50   & 32.04 &  34.61 \\\hline
    \end{tabular}}
    \caption{Tree-level (TL) integrated cross section and that including radiative (rad) corrections for $\nu_\mu-\text{Ar}$ scattering for different neutrino energies.}
    \label{tab:cs-comparison}
\end{table}

\section{Neutrino Charge Radii determination}
\label{app:NCR}
\begin{figure*}[t!]
    \centering
    \includegraphics[width=0.49\linewidth]{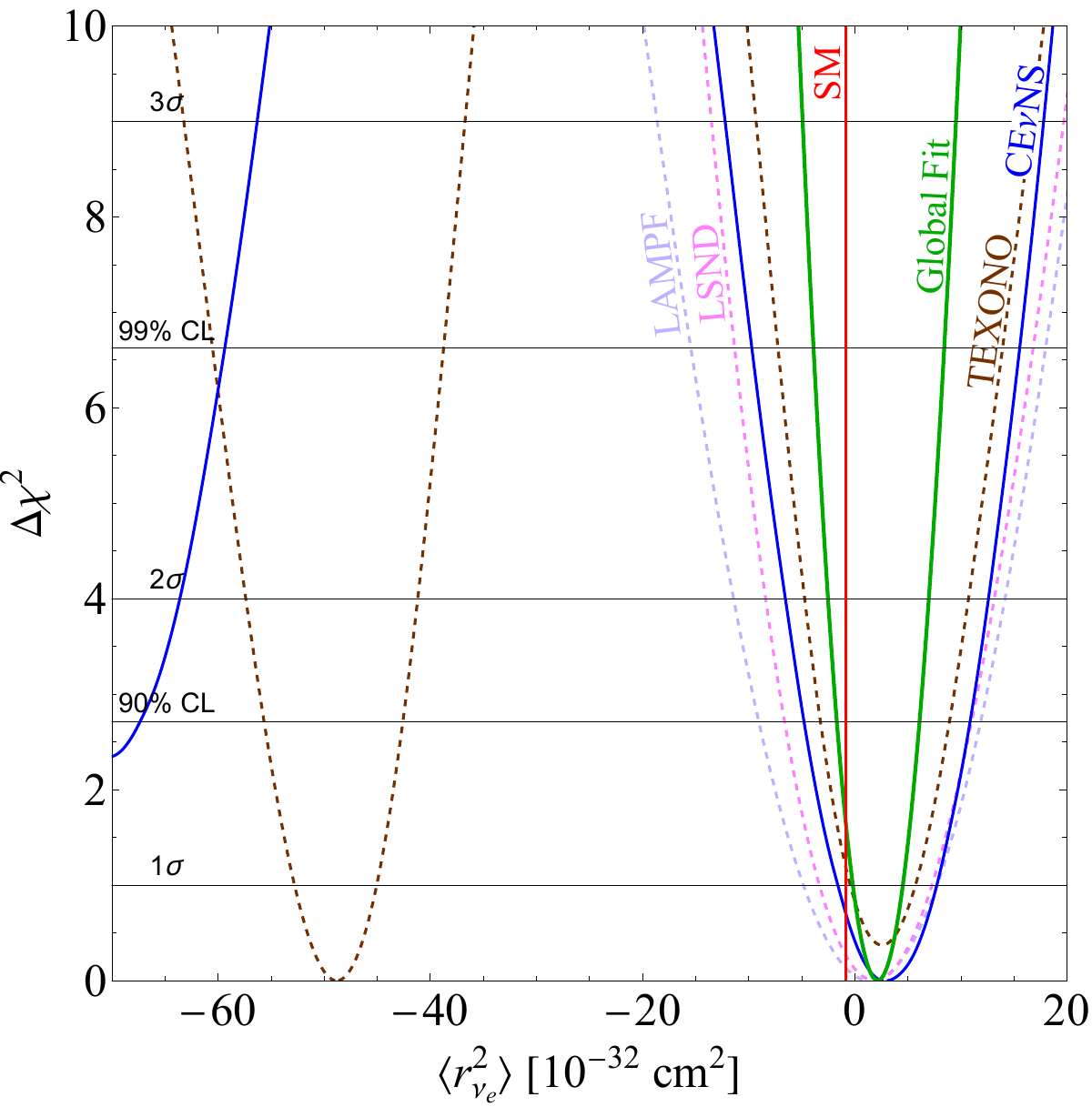}
    \includegraphics[width=0.49\linewidth]{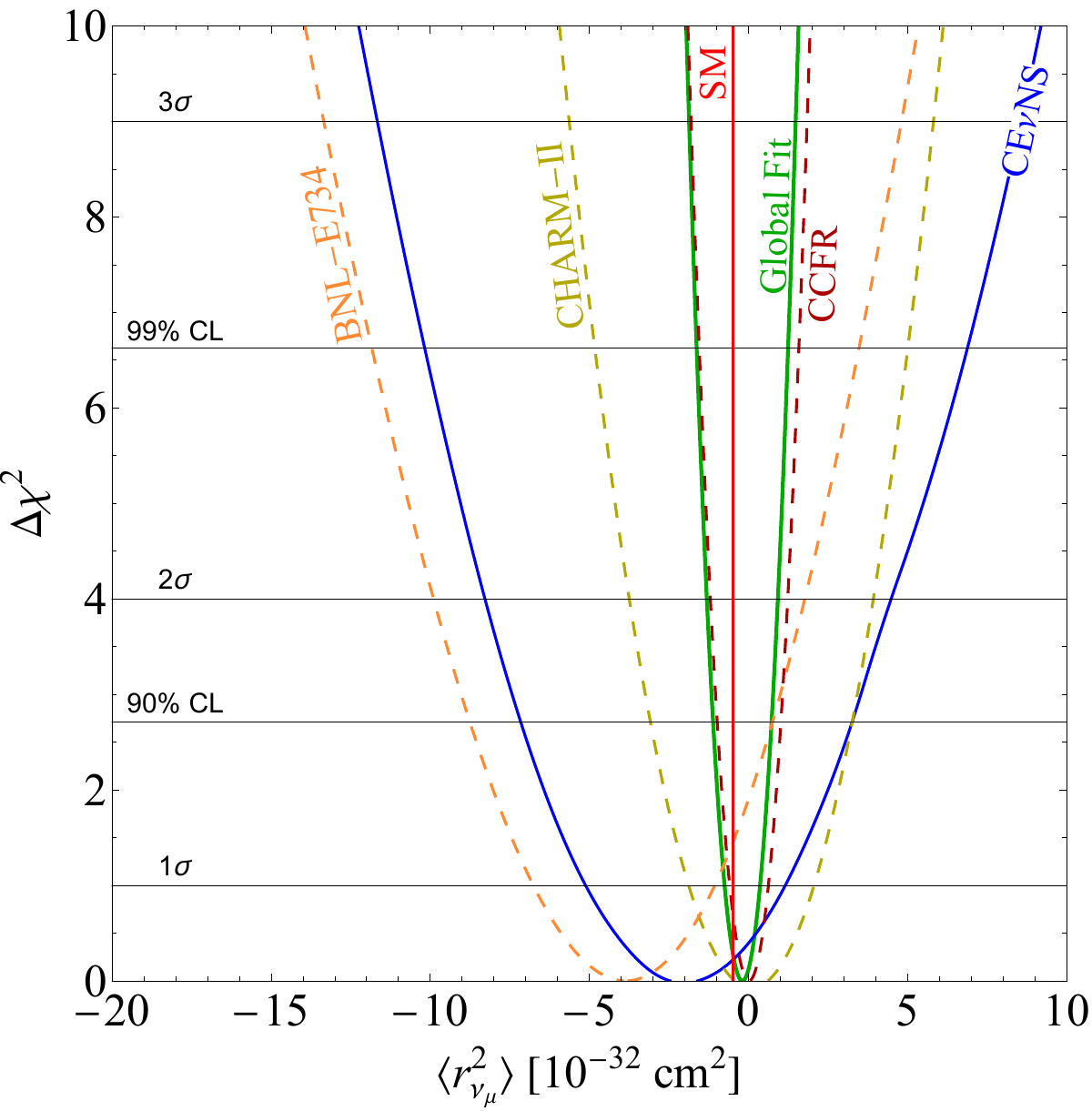}
    \caption{Marginal $\Delta\chi^2$'s of the electron (left) and muon (right) neutrino charge radii obtained from the analysis of $\nu-e$ (dashed curves) and \cenns (solid blue curves) data, along with their combination (green curves). The red vertical lines indicate the SM values for the NCRs.}
    \label{fig:1Deeuu}
\end{figure*}

Here, we provide a more detailed description of the results of the individual determinations of the neutrino charge radii for the experiments included in the global fit. In Fig.~\ref{fig:1Deeuu} we show the marginal $\Delta\chi^2$'s as a function of the electron (left) and muon (right) neutrino charge radii from the analysis of the individual datasets along with their combination. In Tab.~\ref{tab:1Dconstraints}, we report the allowed values at 90\% confidence level. It is worth mentioning that some of the constraints are slightly different from those reported in the literature, see e.g. Refs.~\cite{Giunti:2024gec,Hirsch:2002uv}. However, some of the previous analysis neglected the radiative corrections, and as also shown in Ref.~\cite{Canas:2016vxp}, the obtained results can vary once they are accounted for.
\begin{table}[h]
    \centering
    \renewcommand{\arraystretch}{1.5}
    \begin{tabular}{c|c|c}
        & $\langle r^2_{\nu_e}\rangle\, [10^{-32}\, \mathrm{cm^2}]$&$\langle r^2_{\nu_\mu}\rangle\, [10^{-32}\, \mathrm{cm^2}]$\\\hline
        TEXONO &  [-56,-43]$\cup$[-3.2,8.9] & ---\\\hline
        LSND & [-133,-115]$\cup$[-6.6,11] & ---\\\hline
        LAMPF &  [-133,-113]$\cup$[-9,11.9] & ---\\\hline
        BNL-E734 & --- & [-8.7,0.75] \\\hline
        CHARM-II & --- & [-3.0,3.2] \\\hline
        CCFR & --- & [-0.82,0.92] \\\hline
        CE$\nu$NS & [-73,-67]$\cup$[-4.8,10.9]  & [-58,-50]$\cup$[-7.1,3.2]\\\hline
        Global Fit & [-1.7,6.1] & [-1.1,0.72]\\\hline
    \end{tabular}
    \caption{Allowed values for the NCR at 90\% CL from the fit of the various analysed datasets along with their combination.}
    \label{tab:1Dconstraints}
\end{table}

An important aspect of the obtained constraints is the presence of a degeneracy in the CE$\nu$NS and $\nu$ES cross sections. Indeed, a non-standard value of the neutrino charge radius contribution may lead to an interference with the SM cross section and create a degenerate solution. This is evident in the case of CE$\nu$NS data~\cite{AtzoriCorona:2025ygn,COHERENT:2020ybo,Akimov:2021dab,Ackermann:2025obx,TEXONO:2024vfk,nuGeN:2025mla}, for which the results always present two minima in both the electron and muon flavors, one around the SM prediction and one for large negative values.
Similarly to CE$\nu$NS data, TEXONO~\cite{TEXONO:2009knm}, LSND~\cite{LSND:2001akn} and LAMPF~\cite{Allen:1992qe} data allow for two minima. However, the second minimum from LAMPF and LSND is not visible in Fig.~\ref{fig:1Deeuu} (left) as it is found around $\langle r^2_{\nu_e}\rangle\simeq -120\times 10^{-32}\,\mathrm{cm^2}$ for both datasets. Instead, the degenerate minimum obtained from TEXONO data is visible. The different position of the degenerate solution of TEXONO with respect to the ones of LSND and LAMPF's is due to the different cross section for neutrinos and anti-neutrinos. From the figure, it is clear that CE$\nu$NS finds the degenerate solution for yet other values, due to the combination of proton and neutron couplings in the cross section. Moreover, the neutrino charge radius form factor has a different impact on different datasets, given that they are performed at different experimental energy scales.
The situation seems different for the muonic flavor in Fig.~\ref{fig:1Deeuu} (right). In fact, the $\nu_\mu (\overline{\nu}_\mu)-e$ data show only one minimum, which lies in proximity of the SM prediction. However, this comes from the presence of both $\nu$ and $\overline{\nu}$ events in the BNL-E734~\cite{Ahrens:1990fp}, CHARM-II~\cite{CHARM-II:1994dzw,CHARM-II:1994aeb} and CCFR~\cite{CCFR:1997zzq} datasets, which selects the SM region alone. Clearly, by combining data at different energies, which involve both neutrinos and antineutrinos, it is possible to eliminate the degenerate solutions.
Nevertheless, it is crucial to allow for non-standard values of the NCRs when extracting their measurements, given that data may still prefer the degenerate solution over the SM one. This happens, for example, in the case of TEXONO data, as visible in Fig.~\ref{fig:1Deeuu} (left). The best fit value of the analysis is found to be $\langle r^2_{\nu_e}\rangle\simeq-50\times 10^{-32}\, \mathrm{cm}^2$. If one restricts the fit around the SM prediction, the resulting allowed parameter space would be incorrect.

\section{Momentum dependent correction to $\nu$-e neutral current couplings}
\label{app:gVgA}
Accordingly to Eq.~(\ref{eq:appgv}), the vector coupling between neutrinos and electrons depends on the weak mixing angle. We already discussed the momentum dependence of the neutrino charge radius radiative contribution, which enters the vector coupling. However, also the neutral current flavor independent vector coupling, $\tilde{g}_V^{\nu\, e}$, depends on the momentum transfer due to the running of the weak mixing angle, which can be evaluated in the SM~\cite{Erler:2017knj}.
\begin{figure*}[t!]
    \centering
    \includegraphics[width=0.49\linewidth]{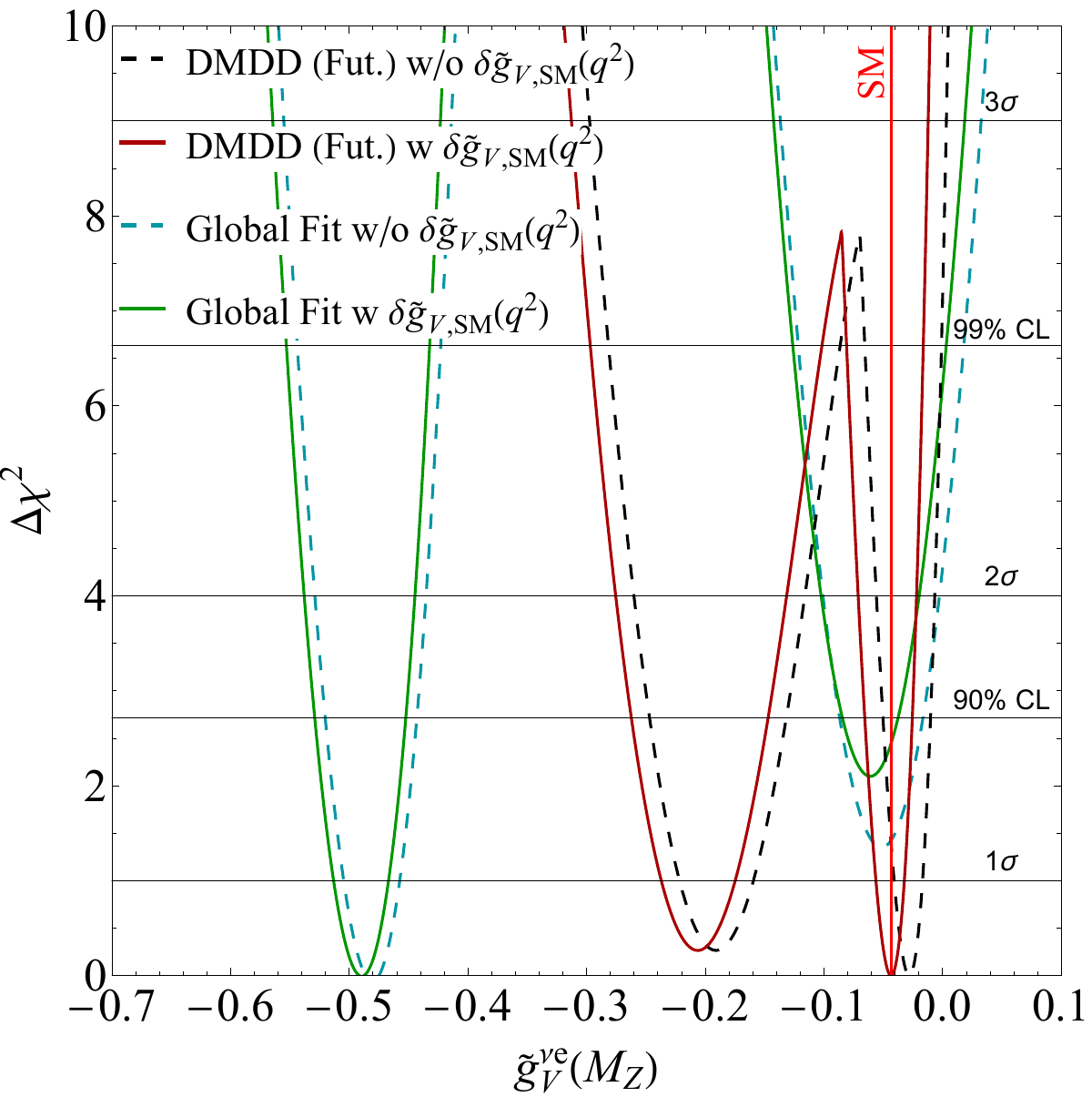}
    \includegraphics[width=0.49\linewidth]{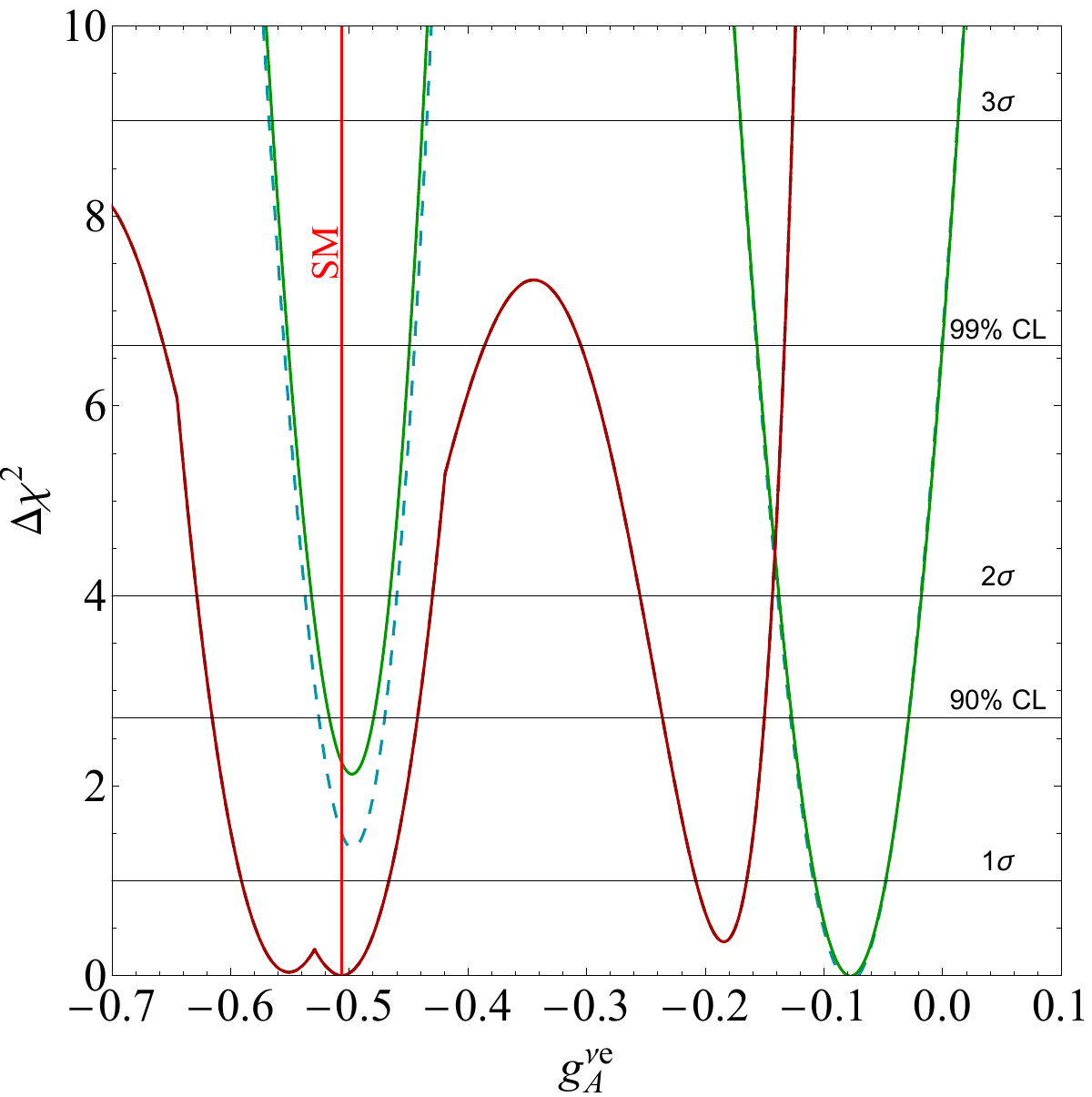}
    \caption{Marginal $\Delta\chi^2$'s of the vector (left) and axial-vector (right) couplings obtained from our global analysis of $\nu$ES data and for the sensitivity from a future DMDD including (solid) or not (dashed) the correction due to the momentum transfer. The red vertical lines indicate the SM values for the couplings.}
    \label{fig:1DgVgA}
\end{figure*}
Given that the data analysed in the global fit have been collected at diverse energy scales, it is clear that $\tilde{g}_V^{\nu\, e}$ entering the scattering cross section is not common at all scales. In order to correct for the effect of the running of the weak mixing angle on $\tilde{g}_V^{\nu\, e}$, we have introduced a term, $\delta \tilde{g}_{V,\, \mathrm{SM}}$, in the definition of the coupling
\begin{equation}
    \tilde{g}_V^{\nu\,e}(q^2)=\tilde{g}_V^{\nu\,e}(M_Z)+\delta\tilde{g}_{V,\, \mathrm{SM}}(q^2)\, .
\end{equation}
Indeed, $\delta \tilde{g}_{V,\, \mathrm{SM}}$ has been defined as the difference between the coupling constant at a certain energy scale $q^2$ with respect to its value at $M_Z$
\begin{align}
    \delta \tilde{g}_{V,\, \mathrm{SM}}(q^2)&=\tilde{g}_{V,\, \mathrm{SM}}^{\nu\,e}(q^2)-\tilde{g}_{V,\, \mathrm{SM}}^{\nu\,e}(M_Z)\, =\nonumber\\
    &=2\rho\left[\sin^2\theta_W(q^2)-s^2_Z\right]\, .
\end{align}
In this way, the coupling at a generic scale $q^2$ can be written in terms of its value at $M_Z$, under the weak assumption that the effect of the weak mixing angle running would not be significantly affected by a deviation of the value of the neutral current coupling at the $Z$ mass scale.
The correction $\delta\tilde{g}_{V,\, \mathrm{SM}}^{\nu\,e}(q^2)$ has a larger impact for low energy data, for which it can reach values of about 50\% of $\tilde{g}_{V,\, \mathrm{SM}}^{\nu\,e}(q^2=0)$.

The effect on the global analysis is shown in Fig.~\ref{fig:1DgVgA}, where we show the marginal $\Delta\chi^2$ of the vector (left) and axial-vector (right) couplings obtained compared to the sensitivity for a future DMDD detector~\cite{Giunti:2023yha}. We compare the results obtained including or not the $\delta\tilde{g}_{V\, \mathrm{SM}}(q^2)$ correction. The constraints on the axial-vector (right) coupling are only marginally affected, while the vector (left) ones present a shift towards smaller values. Future DMDD experiments show the capability of selecting the SM solution over the degenerate one. For such precision levels, it will become crucial to properly account for the effect of the running of the weak mixing angle to avoid misinterpretation of the measurements, as shown by the difference between the solid and dashed lines in Fig.~\ref{fig:1DgVgA}~(left).

\bibliographystyle{apsrev4-2}
\bibliography{supplemental}